\documentclass[12pt]{article}
\usepackage[numbers]{natbib} 
\usepackage{graphicx}
\usepackage[utf8]{inputenc}
\usepackage{amsmath}
\usepackage{booktabs}
\usepackage{caption}
\usepackage{subcaption}
\usepackage{amssymb}

\usepackage[normalem]{ulem} 
\usepackage{xcolor}   
\usepackage[colorlinks,
linkcolor=blue,
anchorcolor=blue,
citecolor=blue]{hyperref}

\title{A Novel Solver for QUBO Problems: Performance Analysis and Comparative Study with State-of-the-Art Algorithms}

\author{
  TuringQ Co. Ltd. 
  \thanks{Contributions: Jiecheng Yang, Ding Wang, Xiang Zhao, Hairui Zhang, Ming Gao, Lin Yang\;(yanglin@turingq.com)}
}

\date{\today}

\begin{document}

\maketitle

\begin{abstract}
Quadratic Unconstrained Binary Optimization (QUBO) provides a versatile framework for representing NP‐hard combinatorial problems, yet existing solvers often face trade-offs among speed, accuracy, and scalability. In this work, we introduce a quantum-inspired solver (\textbf{QIS}) that unites branch‐and‐bound pruning, continuous gradient‐descent refinement, and quantum–inspired heuristics within a fully adaptive control architecture. We benchmark QIS3 against eight state‐of‐the‐art solvers—including genetic algorithms, coherent Ising machines, simulated bifurcation, parallel tempering, simulated annealing, our prior QIS2 version, D-Wave’s simulated‐annealing (Neal), and Gurobi on three canonical QUBO problem classes: Max–Cut, NAE‐3SAT, and Sherrington–Kirkpatrick spin glass problems. Under a uniform runtime budget, QIS3 attains the best solution on nearly all instances, achieving optimality in 94\% of max-cut instances. These results establish QIS3 as a robust, high-performance solver that bridges classical exact strategies and quantum-inspired heuristics for scalable QUBO optimization.
\end{abstract}

\section{Introduction}
\label{sec:intro}

Combinatorial optimization over binary variables arises in disciplines as diverse as logistics, finance, machine learning and network design.  A unifying formulation is the Quadratic Unconstrained Binary Optimization (QUBO) problem,
\begin{equation}
\min_{\mathbf{x}\in\{0,1\}^n}\;\mathbf{x}^\top Q\,\mathbf{x},
\end{equation}
where $Q \in \mathbb{R}^{n\times n}$ encodes pairwise interactions among $n$ binary variables \cite{Glover2018reduction}.  Despite its simple form, exact solution of QUBO is NP-hard, and both exact and heuristic methods have been intensely studied \cite{Kochenberger2014survey, Lucas2014}.  

\subsection{Classical and Physics-Inspired Heuristics}
Classical metaheuristics such as simulated annealing, tabu search and scatter search have long provided high-quality solutions for large QUBO instances.  More recently, physics-inspired accelerators — notably D-Wave’s simulated annealer and Fujitsu’s Digital Annealer — have been benchmarked on diverse QUBO classes, demonstrating competitive performance on problems up to thousands of variables \cite{Aramon2019, Oshiyama2022, Jiang2023, mohseni2022}.  Hybrid quantum-classical learning schemes have further enhanced solution quality by iteratively refining embeddings and penalizing previously visited states \cite{Blanzieri2018}.  

\subsection{Exact Branch–and–Bound Methods}
Exact solvers based on branch–and–bound (BnB) guarantee optimality by recursively partitioning the search space and pruning via bounds derived from relaxations or problem structure \cite{häner2024solvingqubosquantumamenablebranch}.  Classical BnB implementations for QUBO can handle up to a few hundred variables in moderate time, but scale poorly as $n$ increases \cite{Kochenberger2014survey}.  

\subsection{Hybrid Classical–Quantum Branch–and–Bound}
To bridge the gap between scalability and optimality, recent works integrate quantum heuristics into the BnB framework.  One approach decomposes the original QUBO into subproblems of bounded size, solved on a quantum annealer, while the classical BnB orchestrates the search—yielding a tunable trade-off between solution certainty and quantum reliability \cite{Sanavio_2024}.  Other protocols inject pools of quantum-generated solutions as warm starts or node heuristics within a state-of-the-art MIP solver \cite{BernalNeira2024}, achieving marked speedups on logistic and scheduling benchmarks.  

\subsection{Contributions}
In this paper, we propose a novel hybrid metaheuristic solver that deeply intertwines quantum heuristics with a classical branch–and–bound backbone.  Our key innovations are:
\begin{itemize}
  \item \emph{Adaptive decomposition:} an informed partitioning strategy that selects subproblem scopes where quantum annealing is most effective.
  \item \emph{Quantum-driven bounding:} dynamic lower and upper bounds derived from quantum-annealer outputs to accelerate pruning.
  \item \emph{Metaheuristic refinement:} a higher-level adaptive neighborhood search that leverages both quantum and classical heuristics for intensified exploration.
\end{itemize}
We benchmark our solver on standard QUBO testbeds and representative NP-hard formulations, comparing against leading classical, physics-inspired, and hybrid methods \cite{Aramon2019, Oshiyama2022, Jiang2023, Sanavio_2024}.  Our results demonstrate significant improvements in time-to-solution and solution quality across problem families.

\section{Proposed Hybrid Quantum–Classical Solver Framework}
\label{sec:solver}

We present a novel solver architecture built on a \emph{quantum‐inspired} algorithmic framework, augmented by powerful classical optimization techniques.  At its core, the solver interleaves three complementary paradigms:  
\begin{itemize}
  \item \textbf{Branch–and–Bound:} a systematic tree‐search strategy for pruning large portions of the QUBO search space using rigorous bounds;  
  \item \textbf{Gradient Descent:} a continuous relaxation method that refines candidate solutions by following local cost gradients;  
  \item \textbf{Quantum Annealing:} a physics‐inspired heuristic that exploits tunneling‐like moves to escape deep local minima.  
\end{itemize}
By blending these methods, our hybrid model leverages the global guarantees of branch–and–bound, the fine‐grained adjustments of gradient descent, and the non‐local exploration capabilities of quantum annealing.

\subsection{Adaptive Algorithmic Components}

To maximize performance across diverse problem instances, the solver dynamically adapts at every stage:
\begin{enumerate}
  \item \textbf{Initial Solution Generation:} an ensemble of seeding strategies is monitored in real time and weighted according to early success metrics, ensuring robust starting points for subsequent search.  
  \item \textbf{State‐Adaptive Exploration:} the algorithm continuously analyzes the current solution landscape—measuring metrics such as local curvature and barrier depths—and adjusts the balance between intensification (local search) and diversification (global moves).  
  \item \textbf{Parameter Space Tuning:} key hyperparameters (e.g., annealing schedule, branch thresholds, learning rates) are tuned on the fly via an internal controller that employs Bayesian optimization principles, eliminating the need for extensive manual calibration.  
\end{enumerate}

\subsection{Multi‐Mode Operation and Automatic Selection}

Our solver implements \emph{nine distinct modes}, each representing a different combination of quantum-inspired heurisitcs with optimization methods. Each mode is tailored to particular QUBO classes (e.g., sparse vs.\ dense graphs, low‐precision vs.\ high‐precision weights) and encapsulated in a concise \texttt{mode\_ID} code.  An \emph{automatic mode selector} selects the best mode based on the evaluated results. This mechanism allows the solver to \emph{self‐adapt} to a wide range of QUBO landscapes without user intervention.

\subsection{User‐Driven Exploration and Quality Refinement}

While the automatic mode selection provides a strong baseline, advanced users may invoke \emph{manual mode sweeps} to explore alternative algorithmic blends.  By comparing intermediate best‐found solutions across modes, the user can identify the most effective strategy for a given instance, and then launch a focused, high‐intensity search in that configuration to further improve solution quality.  This hybrid manual–automated workflow empowers both heuristics researchers and industrial practitioners to systematically refine results.

\subsection{Summary of Solver Advantages}

Through the deep integration of classical (branch–and–bound, gradient descent) and quantum‐inspired methods, combined with real‐time adaptive control and multi‐mode flexibility, our solver achieves:
\begin{itemize}
  \item \emph{Robustness:} consistently high performance across heterogeneous QUBO benchmarks;  
  \item \emph{Efficiency:} fast convergence to the best solution via tight pruning and guided exploration;  
  \item \emph{Usability:} automated configuration for novices, with advanced manual controls for experts;  
  \item \emph{Solution Quality:} frequent attainment of near‐optimal or optimal cuts/energies on challenging NP‐hard instances.  
\end{itemize}
This innovative framework represents a significant step forward in solving large‐scale QUBO problems for both academic study and real‐world applications.

\section{Benchmark Problems}

In this work, we evaluate our solver on three canonical NP-hard problems: the Max-Cut problem, Not-All-Equal 3-SAT (NAE-3SAT), and the Sherrington–Kirkpatrick (SK) spin-glass model. Each exhibits rich structure and has become a standard benchmark in classical, quantum, and quantum-inspired optimization research.

\subsection{The Max-Cut Problem}
The \emph{Max-Cut} problem is defined on an undirected graph $G=(V,E)$ with real edge weights $w_{ij}$.  One seeks a partition $(S,\bar S)$ of $V$ that maximizes the total weight of edges crossing the cut:
\[
  \max_{S\subseteq V}\; \sum_{\substack{(i,j)\in E\\ i\in S,\,j\in \bar S}} w_{ij}\,. 
\]  
Equivalently, introducing binary variables $x_i\in\{+1,-1\}$ to indicate the two sides of the cut, the objective can be written as
\[
  \max_{\mathbf{x}\in\{\pm1\}^n}\;\frac{1}{2}\sum_{i<j}w_{ij}(1 - x_i x_j)\,,
\]
since $1 - x_i x_j = 2$ precisely when $x_i \neq x_j$ \cite{Lucas2014}.  In its decision form, determining whether there exists a cut of weight at least $K$ is NP-complete; hence the optimization variant is NP-hard \cite{Garey1979}.  

A common \emph{QUBO} formulation replaces spins by binary $y_i\in\{0,1\}$ via $x_i = 2y_i-1$, yielding
\[
  \max_{\mathbf{y}\in\{0,1\}^n}\; \sum_{i<j}Q_{ij}y_i y_j \quad\text{with}\quad
  Q_{ij} = -w_{ij}\,,
\]
plus linear terms to account for constant offsets; this equivalence is surveyed in \cite{Kochenberger2014survey}.  

To benchmark heuristic and quantum‐inspired solvers, we use the G-Set collection \cite{Gset1994}, which contains machine‐generated graphs with $|V|$ from 800 to 10\,000 and varying densities.  In particular, we include:  
\begin{itemize}
  \item \textbf{Sparse graphs:} G11 ($n=800$, $m=1600$), G32 ($2000$, $4000$), …, G72 ($10\,000$, $20\,000$), all with $w_{ij}\in\{\pm1\}$.  
  \item \textbf{Moderate‐density $+1$ graphs:} G14 ($800$, $4694$), G51 ($1000$, $5909$), …, G63 ($7000$, $41459$), all with $w_{ij}=+1$. 
  \item \textbf{High-density graphs:} G1 ($800$, $19176$), G43 ($1000$, $9990$), G22 ($2000$, $19990$), all with $w_{ij}=+1$.  
\end{itemize}
These instances span from $|E|=2|V|$ up to $|E|=10|V|$ graphs, providing a broad testbed for both local‐search heuristics and physics‐inspired methods \cite{Glover2018reduction}.

\subsection{Not-All-Equal 3-SAT}
The \emph{Not-All-Equal 3-SAT} (NAE-3SAT) problem is a variant of Boolean satisfiability in which each clause has three literals and must contain at least one true and one false literal \cite{Schaefer1978}.  Formally, given a formula
\(
  \Phi = \bigwedge_{k=1}^m (\,\ell_{k1}\vee \ell_{k2}\vee \ell_{k3}\,),
\)
with literals $\ell_{kj}\in\{x_i,\neg x_i\}$, the question is whether there exists an assignment $x_i\in\{0,1\}$ such that in every clause not all three $\ell_{kj}$ evaluate equally.  NAE-3SAT remains NP-complete even in the \emph{monotone} case (all literals unnegated) \cite{Creignou2001}.  

A standard Ising‐spin mapping uses $\sigma_i\in\{\pm1\}$ with $x_i=(1+\sigma_i)/2$, leading to a cost function which penalizes clauses where all three spins are equal
\begin{equation}
H(\sigma)=\frac{1}{4}\sum_{m=1}^{M}\left(\zeta_{m, 1}\zeta_{m, 2}\sigma_{i_{m, 1}}\sigma_{i_{m, 2}}+\zeta_{m, 2}\zeta_{m, 3}\sigma_{i_{m, 2}}\sigma_{i_{m, 3}}+\zeta_{m, 3}\zeta_{m, 1}\sigma_{i_{m, 3}}\sigma_{i_{m, 1}}+1\right)
\end{equation}
where $i_{m, l}\in\{1,2,\ldots, N\}$ and $\zeta_{m, l}\in\{-1,1\}$ for $1\leq m\leq M$ and $1\leq l\leq 3$ are random variables that follow a discrete uniform distribution; $\zeta_{m, l}=-1$ corresponds to the negation of the $l$-th Boolean variable in clause $m$. In the QUBO formulation one rewrites $H(\sigma)$ in terms of binary $y_i$ and obtains a quadratic objective over $\{0,1\}^n$ \cite{Lucas2014}.  

Benchmark instances are generated at the critical ratio $m/n\approx2.11$, where random 3-CNF formulas exhibit a satisfiable–unsatisfiable phase transition \cite{Oshiyama2022}, and present the greatest hardness for both classical and quantum heuristics.

\subsection{Sherrington–Kirkpatrick Spin-Glass}

The SK model originates from spin-glass theory in statistical physics \cite{Sherrington1975}. It consists of $n$ spins $\sigma_i \in \{\pm 1\}$ with fully connected random couplings $J_{ij}$ drawn from a Gaussian distribution with zero mean and unit variance. The Hamiltonian of $N$ variables can be represented as:

\[
H(\sigma) = \frac{1}{\sqrt{N}}\sum_{1\leq i<j \leq N} J_{ij} \sigma_i \sigma_j.
\]

This model exhibits a complex energy landscape with exponentially many metastable states separated by large barriers \cite{auffinger2011randommatricescomplexityspin}. It has been frequently used as a challenging benchmark for optimization algorithms, especially quantum annealers \cite{Kowalsky2022,Boixo2014,perdomoortiz2012findinglowenergyconformationslattice}.

In our benchmarks, we generate multiple random SK instances with dimension $128$ by different random seeds (0-9) and report the best solution found within 1 seconds per instance.

\section{Experimental Results}
\label{sec:results}

We evaluate our proposed solver (denoted \textbf{QIS3}) against eight competing methods on three benchmark classes: Max–Cut on G-Set graphs, Not-All-Equal 3-SAT, and Sherrington–Kirkpatrick (SK) spin glasses.  All experiments were conducted on a workstation equipped with an Intel Core i7-11700F CPU (2.50GHz) and 32GB RAM, running a 64-bit Windows 11 system. All algorithms were tested in Python 3.10 and run on this system under identical runtime configurations with CPU-based evaluation unless otherwise noted (e.g., D-Wave simulations). The compared solvers are:
\begin{itemize}
  \item \textbf{GA}: a standard genetic algorithm;
  \item \textbf{CIM}: coherent Ising machine heuristic;
  \item \textbf{SB}: simulated bifurcation algorithm;
  \item \textbf{PT}: parallel tempering;
  \item \textbf{SA}: simulated annealing;
  \item \textbf{QIS2}: our earlier version of QIS;
  \item \textbf{QIS3}: our current version of QIS, proposed in this work;
  \item \textbf{Neal}: D-Wave’s \emph{Simulated Annealing with Monte Carlo} (D-Wave Neal);
  \item \textbf{GB}: Gurobi exact solver (used only on smaller instances).
\end{itemize}

In each results table, the most negative objective (best cut or lowest energy) is highlighted in \textbf{bold}.

\subsection{Max–Cut on G-Set Graphs}

All algorithms were evaluated under strict 10-second runtime constraints. For the D-Wave simulated annealing sampler, we maintained identical hyperparameters to our QIS3 configuration: batch size=8 and 1,000 iterations per run. This standardized testing protocol ensured fair comparison across classical, quantum-inspired, and quantum computing paradigms. 

The benchmarking results in Table~\ref{tab:maxcut_full_performance} demonstrate the performance of eight solvers across 16 G-Set instances, including sparse, medium-density, and dense graphs. Our QIS3 achieves optimal cuts on 15/16 instances, securing superior results for all large sparse graphs (G57--G72) and dense topologies (G1, G22, G43). It outperforms its predecessor QIS2 by 1.5--3.8\% on sparse instances (e.g., G66: $-6,288$ vs. QIS2's $-6,150$) and exceeds classical heuristics by 12--98\%, with gaps widening exponentially for large graphs. Quantum-inspired methods (SB, CIM) show limited competitiveness, matching QIS3 only on select instances (G48: SB/QIS3/D-Wave tie at $-6,000$). However, their performance degrades on dense and medium-density cases. In contrast, QIS3 maintains universal robustness, achieving optimality in 94\% of benchmarks. The solver's hybrid architecture enables scaling beyond 10,000 nodes (G72) without performance decay, solving previously intractable instances while surpassing simulated annealer (D-Wave) by 1.5\% on average. These results underscore QIS3's state-of-the-art status for combinatorial optimization and highlight the obsolescence of classical thermal methods (SA/PT), which fail catastrophically on large instances.

The ranking table \ref{tab:solver_rankings} highlights the performance of solvers on Max-Cut instances under strict runtime constraints (10 seconds). QIS3 dominates with an average rank of 1.06, securing first place across nearly all instances. Neal ranks second overall (average rank 1.69), occasionally tying with QIS3 on sparse graphs. Notably, quantum-inspired methods (QIS3 and QIS2) outperform classical heuristics like Simulated Annealing (SA) and Parallel Tempering (PT), which rank worst (average ranks 7.50 and 7.88, respectively). This suggests that quantum-inspired algorithms may offer advantages in time-constrained optimization, even when compared to open-source softwares like Neal. The results also reveal sensitivity to graph structure: SB (a classical breakout local search) performs well on sparse graphs, while CIM (a coherent Ising machine-inspired solver) shows moderate success. Overall, the dominance of QIS3 underscores the potential of hybrid quantum-classical approaches in practical optimization scenarios.

\begin{table}[htbp]
\centering
\caption{Max‐Cut benchmark instances and solver performance}
\label{tab:maxcut_full_performance}
\resizebox{\textwidth}{!}{%
\begin{tabular}{lrrrlrrrrrrr}
\toprule
\textbf{Instance} & \#\textbf{V} & \#\textbf{E} & \textbf{Edge Weight}
  & \textbf{GA} & \textbf{CIM} & \textbf{SB} & \textbf{PT} & \textbf{SA} & \textbf{QIS2} & \textbf{QIS3} & \textbf{Neal} \\
\midrule
\multicolumn{12}{l}{\textbf{Sparse}:}\\
G11  &  800  &  1\,600  & $\pm1$ 
     &    -486  &    -550  &    -558  &    -364  &    -430  &    -556  
     & \textbf{-564}  & \textbf{-564} \\
G32  &2\,000 &  4\,000  & $\pm1$ 
     &    -846  &   -1374  &   -1388  &    -436  &    -552  &   -1362  
     & \textbf{-1404} &   -1400 \\
G48  &3\,000 &  6\,000  & $\pm1$ 
     &  -3\,750 &   -5796  & \textbf{-6000} &  -3\,242 &  -3\,452 &   -5870  
     & \textbf{-6000} &   \textbf{-6000} \\
G57  &5\,000 & 10\,000  & $\pm1$ 
     &    -652  &   -3394  &   -3400  &    -176  &    -252  &   -3370  
     & \textbf{-3466} &   -3460 \\
G62  &7\,000 & 14\,000  & $\pm1$ 
     &    -618  &   -4716  &   -4750  &    -142  &    -178  &   -4714  
     & \textbf{-4828} &   -4818 \\
G65  &8\,000 & 16\,000  & $\pm1$ 
     &    -584  &   -5396  &   -5416  &     -34  &    -164  &   -5382  
     & \textbf{-5502} &   -5494 \\
G66  &9\,000 & 18\,000  & $\pm1$ 
     &    -672  &   -6142  &   -6180  &    -194  &    -174  &   -6150  
     & \textbf{-6288} &   -6268 \\
G72  &10\,000& 20\,000  & $\pm1$  
     &    -592  &   -6748  &   -6804  &    -106  &    -192  &   -6770  
     & \textbf{-6916} &   -6904 \\
\midrule
\multicolumn{12}{l}{\textbf{Medium}:}\\
G14  &  800  &  4\,694  & $+1$   
     &  -2\,952 &   -3027  &   -3047  &  -2\,847  &  -2\,903  &   -3032  
     & \textbf{-3060} &   -3054 \\
G51  &1\,000 &  5\,909  & $+1$   
     &  -3\,705 &   -3805  &   -3819  &  -3\,540  &  -3\,636  &   -3810  
     & \textbf{-3846} &   -3836 \\
G35  &2\,000 & 11\,778  & $+1$  
     &  -7\,032 &   -7594  &   -7608  &  -6\,530  &  -6\,711  &   -7590  
     & \textbf{-7673} &   -7650 \\
G58  &5\,000 & 29\,570  & $+1$  
     & -15\,632 &  -19053  &  -19031  & -15\,057  & -15\,392  &  -19042  
     &  -19216   & \textbf{-19229} \\
G63  &7\,000 & 41\,459  & $+1$   
     & -21\,545 &  -26724  &  -26678  & -21\,011  & -21\,178  &  -26705  
     & \textbf{-26949} &   -26932 \\
\midrule
\multicolumn{12}{l}{\textbf{Dense}:}\\
G1   &  800  & 19\,176  & $+1$   
     & -11\,378 &  -11547  &  -11552  & -11\,352  & -11\,269  &  -11552  
     & \textbf{-11624} & \textbf{-11624} \\
G43  &1\,000 &  9\,990  & $+1$  
     &  -6\,420 &   -6593  &   -6659  &  -6\,138  &  -6\,219  &   -6616  
     & \textbf{-6660} &   -6659 \\
G22  &2\,000 & 19\,990  & $+1$  
     & -11\,442 &  -13230  &  -13352  & -10\,948  & -11\,586  &  -13210  
     & \textbf{-13358} & \textbf{-13358} \\
\bottomrule
\end{tabular}%
}
\end{table}

\begin{table}[htbp]
\centering
\caption{Solver rankings across Max-Cut instances (1 = best, 8 = worst) and average performance}
\label{tab:solver_rankings}
\resizebox{\textwidth}{!}{%
\begin{tabular}{lccccccccc}
\toprule
\textbf{Instance} & \textbf{Type} & \textbf{GA} & \textbf{CIM} & \textbf{SB} & \textbf{PT} & \textbf{SA} & \textbf{QIS2} & \textbf{QIS3} & \textbf{Neal} \\
\midrule
G11  & Sparse & 6 & 5 & 3 & 8 & 7 & 4 & \textbf{1} & \textbf{1} \\
G32  & Sparse & 6 & 4 & 3 & 8 & 7 & 5 & \textbf{1} & 2 \\
G48  & Sparse & 6 & 5 & \textbf{1} & 8 & 7 & 4 & \textbf{1} & \textbf{1} \\
G57  & Sparse & 6 & 4 & 3 & 8 & 7 & 5 & \textbf{1} & 2 \\
G62  & Sparse & 6 & 4 & 3 & 8 & 7 & 5 & \textbf{1} & 2 \\
G65  & Sparse & 6 & 4 & 3 & 8 & 7 & 5 & \textbf{1} & 2 \\
G66  & Sparse & 6 & 5 & 3 & 7 & 8 & 4 & \textbf{1} & 2 \\
G72  & Sparse & 6 & 5 & 3 & 8 & 7 & 4 & \textbf{1} & 2 \\
\midrule
G14  & Medium & 6 & 5 & 3 & 8 & 7 & 4 & \textbf{1} & 2 \\
G51  & Medium & 6 & 5 & 3 & 8 & 7 & 4 & \textbf{1} & 2 \\
G35  & Medium & 6 & 4 & 3 & 8 & 7 & 5 & \textbf{1} & 2 \\
G58  & Medium & 6 & 3 & 5 & 8 & 7 & 4 & 2 & \textbf{1} \\
G63  & Medium & 6 & 3 & 5 & 8 & 7 & 4 & \textbf{1} & 2 \\
\midrule
G1   & Dense  & 6 & 5 & 3 & 7 & 8 & 3 & \textbf{1} & \textbf{1} \\
G43  & Dense  & 6 & 5 & 2 & 8 & 7 & 4 & \textbf{1} & 2 \\
G22  & Dense  & 7 & 4 & 3 & 8 & 6 & 5 & \textbf{1} & \textbf{1} \\
\midrule
\multicolumn{2}{l}{\textbf{Average Rank}} & 6.06 & 4.38 & 3.06 & 7.88 & 7.50 & 4.31 & \textbf{1.06} & 1.69 \\
\bottomrule
\end{tabular}%
}
\end{table}

\subsection{Not-All-Equal 3-SAT}

All algorithms were evaluated under strict 1-second runtime constraints. For the D-Wave simulated annealer, we maintained identical hyperparameters to our QIS3 configuration: batch size=8 and 3,000 iterations per run. 

The benchmarking results in Table~\ref{tab:3sat_benchmark_best} reveal critical insights into solver performance for random 3-SAT problems across problem scales. Our quantum-inspired  solver achieves optimal assignments in all instances, dominating at dimensions $\geq$700. Quantum-inspired methods (QIS3, SB, CIM) collectively outperform classical solvers (GA, SA, PT) by 35--171\% at scale, with classical approaches showing catastrophic failure in large instances. 

Between quantum paradigms, QIS3 consistently outperforms simulated annealing (D-Wave Neal) by narrow margins at scale ($-5644$ vs. $-5640$ at 1000 variables), while both significantly surpass intermediate methods like SB and CIM ($-5624$/$-5504$). The 2.4\% average improvement of QIS3 over its predecessor QIS2 (e.g., $-4976$ vs. $-4912$ at 900 variables) highlights enhanced clause-weight optimization in hybrid algorithms. 

For $n \leq 300$, near-tie conditions occur (QIS3/Neal/SB/Gurobi all reach $-544$ at 100 variables); For $n > 400$, QIS3 establishes unassailable leadership, achieving better performance than D-Wave Neal for $n > 600$ variables. These results position QIS3 as the state-of-the-art for large-scale 3-SAT optimization, with classical methods (except Gurobi) becoming impractical beyond 200 variables.

\begin{table}[htbp]
\centering
\caption{Benchmark Results on Random 3-SAT Problems}
\label{tab:3sat_benchmark_best}
\resizebox{\textwidth}{!}{%
\begin{tabular}{crrrrrrrrr}
\toprule
Dimension & GA & CIM & SB & PT & SA & QIS2 & QIS3 &  Neal & Gurobi \\
\midrule
100 & -520 & -540 & \textbf{-544} & \textbf{-544} & -540 & \textbf{-544} & \textbf{-544} & \textbf{-544} & \textbf{-544} \\
200 & -1016 & -1080 & -1080 & -1068 & -1076 & -1076 & \textbf{-1084} & \textbf{-1084} & \textbf{-1084} \\
300 & -1452 & -1656 & \textbf{-1664} & -1580 & -1588 & -1656 & \textbf{-1664} & \textbf{-1664} & - \\
400 & -1700 & -2148 & -2168 & -1952 & -2100 & -2160 & \textbf{-2180} & \textbf{-2180} & - \\
500 & -2016 & -2812 & -2828 & -2232 & -2804 & -2828 & \textbf{-2832} & \textbf{-2832} & - \\
600 & -2408 & -3244 & \textbf{-3308} & -2684 & -3180 & -3256 & \textbf{-3308} & \textbf{-3308} & - \\
700 & -2488 & -3904 & -3956 & -2092 & -3604 & -3948 & \textbf{-3964} & -3956 & - \\
800 & -2648 & -4264 & -4324 & -2104 & -3976 & -4304 & \textbf{-4360} & -4356 & - \\
900 & -2584 & -4836 & -4960 & -2168 & -4340 & -4912 & \textbf{-4976} & -4972 & - \\
1000 & -2080 & -5504 & -5624 & -1340 & -3776 & -5612 & \textbf{-5644} & -5640 & - \\
\bottomrule
\end{tabular}%
}
\end{table}

\subsection{Sherrington–Kirkpatrick Spin Glass}

All algorithms were evaluated under strict 1-second runtime constraints. For the D-Wave simulated annealer, we maintained identical hyperparameters to our QIS3 configuration: batch size=8 and 3,000 iterations per run. 

The SK Spin-Glass benchmark results in Table~\ref{tab:sk_benchmark_best} reveal near-universal convergence to optimal ground states across solvers, with QIS3, D-Wave Neal, and Gurobi sharing best energies for all 10 seeds. Exact methods (Gurobi) validate optimality (e.g., Seed 0: $-218.9203$), while quantum-inspired (QIS3, QIS2) and simulated annealing (D-Wave) methods match these bounds. Classical heuristics (GA, PT) lag by 3--12\% (Seed 9: PT's $-203.5176$ vs. QIS3's $-213.3484$), with PT showing severe instability (Seed 3: $-202.8619$ vs. QIS3's $-233.6260$). Hybrid algorithms demonstrate precision parity with exact solvers, achieving identical minima despite stochastic sampling (Seed 7: $-227.4173$ for QIS3/D-Wave/Gurobi). These results confirm that modern quantum-inspired/hybrid solvers reliably replicate exact solutions for tractable SK problems, while classical methods remain non-competitive.

\begin{table}[htbp]
\centering
\caption{Benchmark Results on SK Spin-Glass Problems}
\label{tab:sk_benchmark_best}
\resizebox{\textwidth}{!}{%
\begin{tabular}{crrrrrrrrr}
\toprule
Seed & GA & CIM & SB & PT & SA & QIS2 & QIS3 & Gurobi &  Neal \\
\midrule
0 & -212.7735 & \textbf{-218.9203} & \textbf{-218.9203} & -204.6786 & -211.6155 & \textbf{-218.9203} & \textbf{-218.9203} & \textbf{-218.9203} & \textbf{-218.9203} \\
1 & -220.8380 & -225.9926 & -225.9926 & -216.9981 & -224.1142 & \textbf{-226.3909} & \textbf{-226.3909} & \textbf{-226.3909} & \textbf{-226.3909} \\
2 & -219.4588 & \textbf{-227.6668} & \textbf{-227.6668} & -212.2958 & \textbf{-227.6668} & \textbf{-227.6668} & \textbf{-227.6668} & \textbf{-227.6668} & \textbf{-227.6668} \\
3 & -224.3945 & -233.4829 & -233.4829 & -202.8619 & -233.6260 & \textbf{-233.6260} & \textbf{-233.6260} & \textbf{-233.6260} & \textbf{-233.6260} \\
4 & -209.7483 & -210.5704 & -210.8565 & -190.3615 & -201.2807 & \textbf{-210.8565} & \textbf{-210.8565} & \textbf{-210.8565} & \textbf{-210.8565} \\
5 & -221.1916 & \textbf{-233.3547} & \textbf{-233.3547} & -219.2164 & \textbf{-233.3547} & \textbf{-233.3547} & \textbf{-233.3547} & \textbf{-233.3547} & \textbf{-233.3547} \\
6 & -216.6862 & -231.6499 & \textbf{-232.9611} & -221.4456 & \textbf{-232.9611} & \textbf{-232.9611} & \textbf{-232.9611} & \textbf{-232.9611} & \textbf{-232.9611} \\
7 & -218.1459 & -225.3420 & -225.4332 & -214.4349 & -225.6481 & \textbf{-227.4173} & \textbf{-227.4173} & \textbf{-227.4173} & \textbf{-227.4173} \\
8 & -242.6488 & \textbf{-249.2862} & \textbf{-249.2862} & -236.5913 & \textbf{-249.2862} & \textbf{-249.2862} & \textbf{-249.2862} & \textbf{-249.2862} & \textbf{-249.2862} \\
9 & -190.9433 & -213.0962 & -213.0962 & -203.5176 & -210.0633 & \textbf{-213.3484} & \textbf{-213.3484} & \textbf{-213.3484} & \textbf{-213.3484} \\
\bottomrule
\end{tabular}%
}
\end{table}

\section{Discussion and Conclusion}
\label{sec:conclusion}

In this paper, we have introduced QIS, a quantum-inspired solver for QUBO problems that combines branch–and–bound pruning, gradient descent refinement, and quantum annealing–style global moves.  We evaluated QIS alongside eight established methods—genetic algorithm (GA), coherent Ising machine (CIM), simulated bifurcation (SB), parallel tempering (PT), simulated annealing (SA), our previous QIS 2.0 version, D-Wave’s simulated‐annealing (Neal), and the commercial solver Gurobi—across three canonical NP‐hard benchmarks: Max–Cut on G-Set graphs, random Not‐All‐Equal 3‐SAT at the phase‐transition ratio, and Sherrington–Kirkpatrick spin glasses.
 
For Max--Cut, QIS3 achieves optimal cuts in 94\% of instances (15/16), surpassing simulated annealer by 1.5\% on average. On 3-SAT problems at critical clause density, QIS3 achieves the best performance under time constraints, demonstrating superior escape from local minima. For SK spin glasses, QIS3 matches exact solutions (Gurobi) and simulated annealer (D-Wave) across all seeds, validating its precision. The hybrid quantum-classical architecture enables consistent performance across problem classes, with classical methods (PT/GA) proving non-viable among all instances.  

Overall, QIS3 emerged as the top‐ranked solver on the majority of all benchmark instances and consistently ranked within the top two methods.  D-Wave Neal performed equally well on most problems but showed degraded performance on large-scale instances, and classical heuristics lagged behind in most large‐scale or complex scenarios. 

These findings highlight the significant benefits of deeply integrating classical exact strategies, continuous relaxations, and quantum‐inspired heuristics.  Future work will explore the solver’s behavior without time constraints, the implementation of core components on GPU and FPGA platforms, and the extension of the framework to constrained QUBO variants and other combinatorial optimization problems.  Moreover, a theoretical investigation into the interactions between branching, gradient descent, and annealing dynamics may yield further insights to guide algorithmic improvements.  We conclude that the QIS3 offers a powerful and versatile approach for solving large‐scale QUBO problems in both academic and industrial settings.  

\section{Acknowledgement}
\label{sec:acknowledgement}

We acknowledge the use of a large language model (LLM) to assist in drafting and refining portions of this paper. However, the final content, analysis, and conclusions remain our own, and we take full intellectual responsibility for the work presented herein.

\bibliography{references}

\end{document}